\documentclass{jpp}
\usepackage{graphicx}

\usepackage{hyperref}
\usepackage[utf8]{inputenc}
\usepackage[T1]{fontenc}
\usepackage{amsmath}
\usepackage[capitalise]{cleveref}
\usepackage{subfigure}

\newcommand{\bn}{\begin{equation}}
\newcommand{\en}{\end{equation}}

\newcommand{\simgt}{\gtrsim}

\title{Optimised stellarators with a positive radial electric field}

\author{
P. Helander\aff{1}\corresp{\email{per.helander@ipp.mpg.de}}, 
A. G. Goodman\aff{1},
C. D. Beidler\aff{1},
M. Kuczy{\'n}ski\aff{1},
H. M. Smith\aff{1}
}

\affiliation{\aff{1}Max-Planck-Institut f{\"u}r Plasmaphysik, D-17491 Greifswald, Germany}

\begin{document}

\maketitle

\begin{abstract}
We draw attention to an interesting possibility in the design and operation of stellarator fusion reactors, which has hitherto been considered unrealistic under burning-plasma conditions. Thanks to recent advances in stellarator optimisation theory, it appears possible to create a positive (outward-pointing) radial electric field in the plasma core by carefully tailoring the geometry of the magnetic field. This electric field is likely to expel highly charged impurities from the centre of the plasma through neoclassical transport and thus eliminate, or at least mitigate, a long-standing problem in stellarator physics. Further out, the electric field is expected to suddenly change sign from positive to negative, thus creating a region of strongly sheared flow, which could locally suppress turbulent transport and enhance overall energy confinement. 

\end{abstract}

\section{Introduction}

Already at the dawn of the fusion programme, it was recognised that highly charged impurities may accumulate in the core of a magnetically confined plasma \citep{Taylor-1961}. In a hydrogen plasma with a single species of impurity ions with charge $Z \gg 1$, purely classical diffusion due to gyromotion and Coulomb collisions leads to an extremely peaked impurity density profile $n_Z(r)$, where $r$ denotes the minor radius, given by \citep{HS}
	\bn \frac{n_Z(r)}{n_Z(0)} = \left( \frac{n_i(r)}{n_i(0)} \right) ^Z
	\left(\frac{T_i(0)}{T_i(r)} \right) ^{\frac{Z}{2}-1}, 
	\label{classical peaking}
	\en
where $n_i(r)$ and $T_i(r)$ denote the bulk-ion density and temperature profiles. Unless the latter is steep enough, practically all impurities with $Z \gg 1$ will accumulate in the centre of the plasma. 

Classical transport is rarely large enough to be of practical importance, but similarly dire predictions also hold for neoclassical impurity transport \citep{Connor,Hirshman}. In stellarators, this transport is particularly large and acquires a powerful contribution from the radial electric field, which is weighted by the impurity charge $Z$. At low collisionality, the neoclassical particle flux for each particle species $\sigma$ is given by a transport law,
    $$ \Gamma_\sigma = - D_\sigma n_\sigma \left( \frac{d \ln n_\sigma}{dr} - \frac{e_\sigma E_r}{T_\sigma} + \delta_\sigma \frac{d \ln T_\sigma}{dr} \right), $$
where $n_\sigma$ denotes the number density, $e_\sigma$ the charge, and $T_\sigma$ the temperature of the species in question. The diffusivity $D_\sigma$ and the thermodiffusion coefficient $\delta_\sigma$ depend on the collisionality and will be discussed in greater detail below. Unfortunately, the electric field usually points inward, $E_r(r) < 0$, and thus acts to drive impurity ions into the plasma core. To make things even worse, the beneficial effect of a peaked temperature profile displayed by Eq.~(\ref{classical peaking}), so-called temperature screening, which requires $\delta_\sigma < 0$, is often absent in stellarators, though not in all collisionality regimes 
\citep{Helander-2017a,Velasco-2017,Calvo-2018}. Inward impurity transport is frequently observed in stellarator experiments, where a slow but relentless accumulation of impurities can eventually extinguish the plasma by causing excessive radiation losses. Fortunately, stellarator plasmas are often turbulent enough that impurities are ``flushed out'' of the plasma on a time scale faster than that of neoclassical diffusion, but such turbulent transport also degrades energy confinement. 

The prognosis improves dramatically if the radial electric field could be made positive, at least in the centre of the plasma. The neoclassical transport would then be reversed and impurities no longer accumulate there. Such conditions indeed arise in stellarators at low density if the plasma electrons are heated to such an extent that they become substantially hotter than the ions, $T_e > T_i$. As expected theoretically (see below), the radial electric field then changes sign and impurities are expelled from the plasma.\footnote{In LHD, electron roots have also been attained in impure, very-low-density plasmas with $T_i > T_e$  \citep{Fujita_2021}, i.e. under conditions very different from those required in a reactor.} However, in a typical fusion reactor, $T_e \simeq T_i$ since the density needs to be so large and the confinement so good that the energy confinement time exceeds the ion-electron temperature equilibration time. 

It therefore came as a great but pleasant surprise when it was recently found, somewhat serendipitously, that a positive radial electric field spontaneously arose in reactor transport calculations performed for a quasi-isodynamic stellarator \citep{Beidler-2024}. In the present article, we analyse this phenomenon in greater detail and explore the conditions under which the geometry of a stellarator magnetic field is predicted to lead to such behaviour. We also discuss the implications for plasma transport, including the possible appearance of a turbulent transport barrier \citep{Ida}. Most importantly, we show that the appearance of a positive electric field is possible in stellarators optimised for a large number of other favourable properties.

\section{Radial electric field}

In this section, we review the basic scalings of neoclassical and turbulent transport in stellarators, and draw conclusions regarding ambipolarity and the radial electric field. 

\subsection{Neoclassical and turbulent transport}

In stellarators, particle and energy transport are caused by both neoclassical transport and turbulence. The diffusion coefficient associated with the latter is, according to local gyrokinetic theory, of the order of the gyro-Bohm value \citep{Hagan-Frieman,Connor-1988}
	$$ D_{\rm gB} \sim \rho_{\ast i}^2 v_{Ti} a, $$
where $v_{Ti} = \sqrt{2 T_i/m_i}$ is the ion thermal velocity and $\rho_{\ast i} = \rho_i/a$ the gyroradius normalised to the minor radius $a$.\footnote{Here and in the following, we do not always distinguish between the minor radius, the gradient length scale of plasma parameters, and the radius of curvature of the magnetic field lines. For the sake of simplicity, these quantities are treated as comparable and the aspect ratio of the stellarator thus as an ``order unity quantity''.} The magnitude of the neoclassical transport depends sensitively on the geometry of the magnetic field and the collisionality. The electrons are usually in the ``$1/\nu$ collisionality regime'', in which the diffusion coefficient scales as \citep{Galeev,Sagdeev,Ho,Beidler-ICNTS}
	\bn D^{1/\nu}_e \sim \frac{\epsilon_{\rm eff}^{3/2} v_{de}^2}{\nu_e}, 
	\label{De}
	\en
where $v_{de} \sim \tau \rho_{\ast i} v_{Ti}$ denotes the electron drift velocity and $\nu_e$ the electron collision frequency, which is related to that of the ions by 
	$$ \nu_e \sim \frac{M^{1/2} \nu_i}{\tau^{3/2}}, $$
with $\tau = T_e/T_i$ and $M = m_i / m_e$ \citep{HS}. The so-called effective magnetic ripple $\epsilon_{\rm eff}$ depends on the geometry of the magnetic field and measures the net effect of the radial magnetic drift, integrated over all trapped orbits, for this type of transport \citep{Nemov}. It vanishes in an exactly omnigenous field, where, by definition, drift surfaces and flux surfaces coincide. An important goal of stellarator optimisation is to make the ratio of neoclassical to turbulent transport smaller than unity,
	\bn \frac{D^{1/\nu}_e}{D_{\rm gB}} \sim \frac{\epsilon_{\rm eff}^{3/2}\tau^{7/2}}{M^{1/2} \nu_{\ast i}} \lesssim 1, 
	\label{De/DgB}
	\en
where $\nu_\ast = \nu_i a / v_{Ti}$. In practice, $\epsilon_{\rm eff}$ therefore needs to be a few percent or smaller for relevant values of the plasma parameters. That this goal is achievable in a large, low-collisionality stellarator, has recently been demonstrated by experiments in W7-X \citep{Beidler-Nature}.

The ions are usually in a different collisionality regime, the so-called $\sqrt{\nu}$-regime, where the diffusion coeffcient scales as \citep{Galeev,Ho,Calvo-2017}
	\bn D_i^{\sqrt{\nu}} \sim c_{\sqrt{\nu}} \frac{\nu_i^{1/2} \rho_{\ast i}^2 v_{Ti}^2}{\omega_E^{3/2}}
	\sqrt{\ln \left( \frac{\omega_E}{\nu_i} \right)}. 
	\label{sqrt(nu)}
	\en
Here, the coefficient $c_{\sqrt{\nu}}$ encapsulates the effect of the magnetic-field geometry on this type of transport, and is equal to $c_{\sqrt{\nu}} = (b_{10}/\epsilon_t)^2$ in a classical stellarator, where $b_{10}$ measures the lowest-order poloidal harmonic in the variation of the magnetic field strength, and $\epsilon_t$ is the so-called toroidal ripple. We refer the reader to \cite{Beidler-ICNTS} for more details. The denominator contains the ${\bf E} \times {\bf B}$ rotation frequency $\omega_E = |E_r |/ (rB)$, where $E_r$ denotes the radial electric field, $r$ the minor radius, and $B$ the magnetic field. The logarthmic factor, which was relatively recently found by \cite{Calvo-2017}, will be ignored in the following since it is rarely very different from unity. 

Depending on the magnetic-field geometry and the collisionality, regimes other than Eq.~(\ref{sqrt(nu)}) are also possible \citep{Mynick}. In practice, these are not well separated from one another, and an accurate calculation of the transport must therefore be done numerically. A lower bound on the transport is given by the banana-regime coefficient associated with an orbit width proportional to the gyroradius,
	\bn D_i^{\rm ban} \sim c_{\rm ban} \nu_i \rho_i^2, 
	\label{Dban}
	\en
which has a scaling similar to that of classical (Braginskii) transport, tokamak banana transport, and Pfirsch-Schl{\"u}ter transport \citep{HS}. (A similar bound applies to the electrons, which is however extremely small.) An exactly omnigenous stellarator would exhibit such transport \citep{Boozer-1983,HN} and is indeed seen numerically in extremely well optimised configurations \citep{Goodman-2023}. 

\subsection{Ambipolarity}

An important qualitative difference between neoclassical and turbulent transport is the issue of intrinsic ambipolarity. Particle transport is said to be intrinsically (or automatically) ambipolar if the radial electric current, i.e. the sum over all species of the charge-weighted particle fluxes, vanishes for any value of the radial electric field. Stellarator neoclassical transport is never intrinsically ambipolar unless the magnetic field is quasisymmetric \citep{Helander-Simakov,Helander-ROP}. Only for certain specific values of the radial electric field does the radial current vanish. 

In contrast, turbulent transport is intrinsically ambipolar to leading order in the gyroradius expansion according to standard gyrokinetic theory \citep{Sugama,Parra-2009,Helander-ROP}. Any non-ambipolar contribution to the turbulent transport is thus expected to be associated with a diffusion coefficient of at most
	$$D_{\rm turb}^{\rm na} \lesssim \rho_{\ast i} D_{\rm gB}. $$ 
As a result, even if a stellarator has been optimised well enough that Eq.~(\ref{De/DgB}) holds, the nonambipolar part of the transport will usually be dominated by the neoclassical contribution. Only if the geometry of the magnetic field has been optimised to the extent that 
	$$ {D_e^{1/\nu}}\lesssim \rho_\ast {D_{\rm gB}} $$
does the turbulent transport significantly affect the ambipolarity condition. Such extreme optimisation is however unnecessary for energy confinement and is rarely undertaken in practice. As a result, although most of the energy transport in large stellarators such as LHD and W7-X \citep{Beidler-Nature} is caused by plasma turbulence, the radial electric field is nevertheless expected to be determined by neoclassical transport to leading order in the smallness of the gyroradius \citep{Sugama,Helander-Simakov,Helander-ROP}. An exception occurs on small radial scales, where short-wavelength zonal flows can be excited by turbulence \citep{Helander-Simakov}. 

\subsection{Electron and ion roots}

In non-quasisymmetric stellarators, then, ambipolarity is not automatic but is realised only for one or several particular values of the radial electric field. The dependence of the radial current on $E_r$ is nonlinear, and it has long been known that it can have up to three roots \citep{Mynick-Hitchon,Shaing,Hastings-1985,Hastings-1986}. If the ion and electron temperature are comparable, there is usually only the so-called ``ion root'' associated with an inward-pointing electric field, $E_r < 0$. However, if the electrons are relatively hot, $\tau > 1$, an ``electron root'' with $E_r >0$ is possible. A third, intermediate, root is unstable and therefore expected to be irrelevant.\footnote{For this root the radial current $J_r = e( \Gamma_i - \Gamma_e)$ varies inversely with the electric field, $\partial J_r / \partial E_r < 0$. Any small deviation in $E_r$ from the ambipolar value will thus be amplified. }

These theoretical expectations are borne out in practice. Stellarator experiments demonstrate that the radial electric field is usually negative but switches sign in the centre of the plasma if strong heating is selectively applied to the electrons. In a range of several very different stellarators (CHS, LHD, TJ-II, W7-AS), localised electron cyclotron resonance heating (ECRH) resulted in a regime of improved confinement, the so-called ``core electron-root confinement regime,'' associated with postive $E_r$ \citep{Yokoyama}. Above a certain ECRH power, the electron temperature profile became very steep, enabling a relatively high electron temperature to be reached in the core. Similar plasmas have more recently been created in W7-X \citep{Geiger}. In HSX, however, only the ion root has been observed although the electrons are usually considerably hotter than the ions, but this stellarator is quasihelically symmetric and should therefore behave differently due to intrinsic ambipolarity. 

Theoretically, the ion root appears because the electrons are the ``rate-controlling'' species \citep{Ho}. In the absence of an electric field, both the electrons and the ions would be in their respective $1/\nu$ regimes, but the ion transport would then be much larger than the electron transport, violating ambipolarity. A radial electric field must therefore arise to reduce the ion transport to the electron level, which thus sets the overall transport. The electric field makes the ion transport follow the $\sqrt{\nu}$-scaling (\ref{sqrt(nu)}) and adjusts to a value where the latter becomes comparable to Eq.~(\ref{De}). 

For a hydrogen plasma with the electrons in the $1/\nu$-regime and the ions in the $\sqrt{\nu}$-regime, the ambipolarity condition is
    $$ D_e^{1/\nu} \left( \frac{d\ln n}{dr} + \frac{e E_r}{T_e} + \delta_e \frac{d \ln T_e}{dr} \right)  
    = D_i^{\sqrt{\nu}} \left( \frac{d\ln n}{dr} - \frac{e E_r}{T_i} + \delta_i \frac{d \ln T_i}{dr} \right), $$
with $\delta_e = 7/2$ and $\delta_i = 5/4$ \citep{Beidler-Nature}, and  implies a radial electric field given by
    $$ \frac{e E_r}{T_e} = \frac{1 + \eta_i \delta_i - (1 + \eta_e \delta_e) x}{x + \tau} \cdot  \frac{d\ln n}{dr} , $$
where $x =D_e^{1/\nu} / D_i^{\sqrt{\nu}}$ and $\eta_\sigma = d \ln T_\sigma / d \ln n$. For a conventional density profile with $n'(r) < 0$, a positive radial electric field is only realised if
    $$ x > \frac{1 + \eta_i \delta_i }{1 + \eta_e \delta_e}. $$

In order to derive a crude scaling law for plasma parameters admitting an electron root, we take the right-hand-side of this equation to be of order unity and radial scale lengths to be comparable to $a$. The electric field then becomes of order 
	$$ E_r \sim \frac{T_i r}{e a^2}, $$
so that $\omega_E \sim \rho_{\ast i} v_{Ti}/a$. If the logarithmic factor in Eq.~(\ref{sqrt(nu)}) is neglected and $r \sim a$, the condition $x \simgt 1$ leads to the criterion
	\bn\tau \simgt M^{1/7} c_{\sqrt{\nu}}^{2/7} 
	\left( \frac{\nu_{\ast i}}{\epsilon_{\rm eff} \rho_{\ast i}} \right)^{3/7} \equiv \tau_\ast, 
	\label{taustar}
	\en
which should approximately predict the appearance of an electron root. The scalings in this criterion have recently been found to agree with global gyrokinetic simulations \citep{Kuczynski}. Several interesting conclusions can be drawn from this result.

Firstly, it explains why an electron root is at all possible despite the large mass difference between electrons and ions, $M \sim 10^3$. The latter is responsible for the fact that ions have larger neoclassical transport than electrons, and one may be forgiven for wondering whether the reverse can ever be possible. However, the electron-to-ion temperature ratio at which this happens is, according to Eq.~(\ref{taustar}), only proportional to $M^{1/7}$, which is about 3 in a hydrogen plasma. The electrons therefore need not be very much hotter than the ions to enable the appearance of an electron root. 

Secondly, it is interesting to note that, although $\rho_{\ast i}$-scaling in Eq.~(\ref{taustar}) is unfavourable for large stellarators, such devices have relatively good confinment, enabling higher temperatures and lower collisionalities. Thanks to the favourable scaling of Eq.~(\ref{taustar}) with $\nu_{\ast i}$, which enters as $\nu_{\ast i} / \rho_{\ast i}  \propto a^2 n /T_i^{5/2}$,  the electron-to-ion temperature ratio $\tau_\ast$ necessary for an electron root is therefore not necessarily much higher in a stellarator reactor than in present experiments. However, the ion and electron temperatures will almost be equal in a reactor, since the energy confinement time is likely to exceed the Braginskii temperature-equilibration time, making it difficult to attain $\tau > 1$. 

Finally, and most importantly, although electron roots are usually only observed when the electrons are hotter than the ions, Eq.(\ref{taustar}) suggests that an electron root should be possible even if $\tau = 1$. If a stellarator could be optimised to have a sufficiently small ratio $c_{\sqrt{\nu}} / \epsilon_{\rm eff}^{3/2}$, an electron root should be attainable in temperature-equilibrated plasmas. 

These conclusions do not change much if the ion transport is instead given by the lower bound of Eq.~(\ref{Dban}). Then $D_e > D_i$ when
	$$ \tau > \frac{c_{\rm ban}^{2/7}}{\epsilon_{\rm eff}^{3/7}} \; M^{1/7} \nu_{\ast i}^{4/7}, $$
which shares many features with Eq.~(\ref{taustar}). Only the scaling with $\rho_{\ast i}$ is significantly different. 

\section{Optimising for small ion transport}

We turn to the question whether optimisation for an electron root is theoretically possible. The transport coefficient (\ref{sqrt(nu)}) is proportional to $\sqrt{\nu_i}$ due to a peculiar boundary layer in velocity space between trapped and circulating particles \citep{Galeev,Ho,Calvo-2017}. This type of transport is caused by the fact that ions in the boundary layer drift radially whilst undergoing collisional scattering across the trapping boundary. The result is a random walk with a step length proportional to the radial excursion of the orbits between subsequent scatterings. The coefficient $c_{\sqrt{\nu}}$ in Eq.~(\ref{sqrt(nu)}) should therefore be relatively small in a stellarator where particles close to the trapping boundary do not drift much radially between successive bounce points. In contrast, the coefficient $\epsilon_{\rm eff}$ appearing in the electron diffusion coefficient (\ref{De}) depends on the radial drift of {\em all} trapped particles. If the magnetic field could be tailored in such a way that barely trapped particles make smaller radial excursions than more deeply trapped ones, then the ratio $c_{\sqrt{\nu}} / \epsilon_{\rm eff}^{3/2}$ appearing in Eq.~(\ref{taustar}) should become small and the electron root should become relatively easy to attain. In order to understand how this can be achieved, we first recall the general conditions under which radial drifts of trapped particles are small. 

\subsection{Complete omnigenity}

\citet{Hall} introduced the notion of magnetic fields in which the net radial drift of all trapped particles vanishes, and called such fields omnigenous. \citet{Cary} derived a basic criterion for exact omnigenity, which we now rederive. If the magnetic field is written as ${\bf B} = \nabla \psi \times \nabla \alpha$, with $\psi$ denoting the toroidal magnetic flux divided by $2 \pi$ and $\alpha = \theta - \iota \varphi$ in Boozer coordinates, the net drift of a magnetically trapped particle in the $\psi$-direction is equal to \citep{Helander-ROP}
	\bn \Delta \psi = \frac{1}{q} \frac{\partial \mathcal{J}}{\partial \alpha} 
	\label{Delta psi}
	\en
where 
\begin{equation}
    \mathcal{J}(\psi,\alpha,y) \equiv \int mv_\parallel dl = \sqrt{2m\cal{E}} \int\sqrt{1-B(\psi,\alpha,l)/y}\;dl.
    \label{eq:J}
\end{equation}
is the so-called second (or parallel) adiabatic invariant. Here $m$ denotes the mass, $q$ the charge, and $v_\| = \pm v \sqrt{1- B/y}$ the parallel velocity of the particle, where $y = \cal E / \mu$ is the ratio of kinetic energy ${\cal E} = mv^2/2$ to magnetic moment $\mu = m v_\perp^2/2B$, both of which are constant for a particle moving within an equipotential magnetic surface. The integral is taken along the magnetic field between two consecutive bounce points of equal field strength $y$, with the arc length along the field line between these points denoted by $l$. 

It follows that omnigenity is equivalent to the condition
	\bn \frac{\partial}{\partial \alpha} \int \sqrt{y-B(\psi,\alpha,l)} \; {dl} = 0, 
	\label{omnigenity}
	\en
for all values of $y \in [B_{\rm min}(\psi),B_{\rm max}(\psi)]$ between the minimum and maximum field strengths on the magnetic surface in question. Differentiating this equation with respect to $y$ and changing the integration variable from $l$ to $B$ gives
	\bn \frac{\partial}{\partial \alpha} \int_{B_{\rm min}}^y F(\psi,\alpha,B) 
	\frac{dB}{\sqrt{y-B}} = 0, 
	\label{Abel1}
	\en
where the function $F$ is defined by
	$$ F(\psi,\alpha,B) = \sum \left| \frac{\partial B}{\partial l} \right|^{-1}, $$
and the sum is taken over the two bounce points. Equation (\ref{Abel1}) is an Abel integral equation,
	$$ \int_a^y f(x) \frac{dx}{\sqrt{y-x}} = g(y), $$
whose general solution is
	\bn f(y) = \frac{1}{\pi} \frac{d}{dy} \int_a^y g(x) \frac{dx}{\sqrt{y-x}}.
	\label{Abel solution}
	\en
Since the right-hand side of Eq.~(\ref{omnigenity}) vanishes, we must therefore have
	\bn \frac{\partial F}{\partial \alpha} = 0, 
	\label{CS}
	\en
which is the omnigenity criterion derived by \citet{Cary}. As they note, it follows that, for any function $h(\psi,B)$, 
	$$ \frac{\partial}{\partial \alpha} \int_{B(\psi,\alpha,l) \le y} h[\psi,B(\psi,\alpha,l)] dl = 0, $$
for all $y \in [B_{\rm min}(\psi), B_{\rm max}(\psi)]$. For instance, the choice $h=1$ implies that the distance along the field between two consecutive bounce points with field strength $y$ is independent of $\alpha$. In other words, the distance between bounce points is the same for all field lines on the same flux surface. This result is a powerful tool for proving theorems about plasmas confined by omnigenous magnetic fields \citep{HN,Helander-2011,Landreman-Catto}.

\subsection{Partial omnigenity}

In a perfectly omnigenous field, there is neither $1/\nu$-transport nor $\sqrt{\nu}$-transport, i.e. $\epsilon_{\rm eff} = c_{\sqrt{\nu}} = 0$, and the neoclassical transport will therefore be small and tokamak-like. In any actual stellarator, however, the net radial drift never vanishes for all orbits, and there will be some enhancement of the neoclassical transport above the axisymmetric base level. It is therefore of interest to consider how the Cary-Shasharina theorem (\ref{CS}) is modified if the net radial drift (\ref{Delta psi}) vanishes for some, but not all, trapped orbits. 

The simplest case occurs if deeply trapped particles are omnigenous but less deeply trapped ones are not.\footnote{We use the term ``omnigenous'' somewhat unconventionally here, using it to designate particle orbits with no bounce-averaged radial drift. Originally it was introduced in plasma physics to characterise magnetic fields in which all trapped orbits have this property \citep{Hall}, but the word itself is much older. According to the {\em Oxford English Dictionary}, it made its first recorded appearance in the English language in 1650 in the writing of Nathaniel Ward, clergyman and compiler of a law code for Massachusetts. } The integral equation (\ref{omnigenity}) is thus taken to hold for all values of $y$ in the range $B_{\rm min} \le y \le B_1$, where $B_{\rm min} \le B_1 < B_{\rm max}$ and $B_{\rm min}$ is independent of $\alpha$ due to omnigenity \citep{Cary,Landreman-Catto,Helander-ROP}. The solution is then the same as before, Eq.~(\ref{CS}), but it is only valid for $y \in [B_{\rm min},B_1]$ and not for $y > B_1$. The reason is obvious: the deeply trapped particles (i.e. those with $y < B_1$) only sample the magnetic field in the omnigenous part of the trapping well, and are thus unaffected by the lack of omnigenity for more shallowly trapped particles (i.e. those with $y > B_1$).

More interesting, and much more relevant to our present considerations, is the opposite case, in which shallowly trapped particles are taken to be omnigenous but deeply trapped ones are not.\footnote{Strictly speaking, the very most shallowly trapped particles cannot be exactly omnigenous \citep{Cary,Rodríguez_Helander_Goodman_2024}, but we ignore this complication here as it turns out that they can be so to a sufficient approximation.} The question is whether particles that are shallowly trapped can be omnigenous although they must pass through regions where more deeply trapped ones, which are not omnigenous, reside. We thus examine whether Eq.~(\ref{Abel1}) can hold for $y \in [B_1,B_{\rm max}]$ but not for $y < B_1$. In the former region, the following integral equation must then be satisfied,
	$$ \int_{B_1}^y \frac{\partial F(\psi,\alpha,B)}{\partial \alpha} 
	\frac{dB}{\sqrt{y-B}} 
	= - \frac{\partial}{\partial \alpha} 
	\int_{B_{\rm min}(\psi,\alpha)}^{B_1} F(\psi,\alpha,B) \frac{dB}{\sqrt{y-B}},$$
where we note that $B_{\rm min}$ may depend on $\alpha$ since the deeply trapped particles are not omnigenous. According to Eq.~(\ref{Abel solution}), the solution is 
	$$ \frac{\partial F(\psi,\alpha,y)}{\partial \alpha} 
	= - \frac{1}{\pi} \frac{\partial^2}{ \partial y \partial \alpha} 
	\int_{B_1}^y \frac{dx}{\sqrt{y-x}} 
	\int_{B_{\rm min}(\psi,\alpha)}^{B_1} F(\psi,\alpha,B) \frac{dB}{\sqrt{x-B}}, $$
where the two integrals can be interchanged and the one over $x$ performed, 
	$$ \int_{B_1}^y \frac{dx}{\sqrt{(y-x)(x-B)}} = \pi -2 \arctan \sqrt{\frac{B_1 - B}{y-B}}, $$
giving
	\bn \frac{\partial F(\psi,\alpha,y)}{\partial \alpha} 
	= - \frac{1}{\pi\sqrt{y-B_1}} \frac{\partial}{\partial \alpha} \int_{B_{\rm min}(\psi,\alpha)}^{B_1}
	F(\psi,\alpha,B) \frac{\sqrt{B_1 - B}}{y-B} \;{dB}. 
	\label{solution}
	\en
It is straightforward, though slightly tedious, to verify that this function has the property
	$$ \int_{B_1}^y \frac{\partial F(\psi,\alpha,B)}{\partial \alpha} \sqrt{y-B} \; dB
	= - \frac{\partial }{\partial \alpha} \int_{B_{\rm min}}^{B_1} F(\psi,\alpha,x) 
	\left( \sqrt{y-x} - \sqrt{B_1 - x} \right) dx, $$
and thus 
indeed satisfies the integral equation (\ref{Abel1}), but it does {\em not} satisfy Eq.~(\ref{omnigenity}) since with this solution
    $$ \frac{\partial }{\partial \alpha} \int_{B_{\rm min}}^y F(\psi,\alpha,B) \sqrt{y-B} \; dB = \frac{\partial }{\partial \alpha} \int_{B_{\rm min}}^{B_1} F(\psi,\alpha,x) \sqrt{B_1 - x} \; dx.  $$
The right-hand side of this equation vanishes only if particles with $y = B_1$ are omnigenous, which however will not generally be the case in a field that is non-omnigneous for $B(l) < B_1$. Unsurprisingly, if orbits with $y < B_1$ are not omnigenous, those with $y = B_1$ will in general not be so either. 

However, it is possible that orbits with slightly larger values of the bounce-point field strength are omnigenous. Indeed, given the dependence of the magnetic-field strength on $(\psi,\alpha,l)$ for values up to $B_1$, it is possible to extend $B(\psi,\alpha,l)$ to values $[B_1,B_{\rm max}(\psi)]$ in such a way that all particle orbits with $y \ge B_2$ are omnigenous, where $B_2$ is an arbitrarily chosen field strength in the interval $(B_1,B_{\rm max})$. To do so, we merely need to define $F(\psi,\alpha,y)$ for $y \in [B_1,B_2]$ in such a way that 
	$$ \frac{\partial }{\partial \alpha} \int_{B_{\rm min}}^{B_2} 
	F(\psi,\alpha,x) \sqrt{B_2 - x} \; dx =0, $$
which can, for instance, be achieved by choosing
	$$ \frac{\partial F(\psi,\alpha,y)}{\partial \alpha} = - \frac{3}{2} \left(B_2-B_1\right)^{-3/2}
	\frac{\partial}{\partial \alpha} \int_{B_{\rm min}}^{B_1} F(\psi,\alpha,x) \sqrt{B_2 - x} \; dx $$
independently of $y$ throughout the region $y \in [B_1,B_2]$. 

We are thus led to the conclusion that it should be possible to devise magnetic fields in which deeply trapped particles are less well confined than shallowly trapped ones. This should improve the neoclassical confinement of ions insofar as they are in the $\sqrt{\nu}$-regime since this type of transport vanishes if orbits close to the trapped-passing boundary are perfectly omnigenous. According to the estimate (\ref{taustar}), the electron-to-ion temperature ratio necessary for an electron root would then be reduced, and there is no reason why it could not be as low as unity or smaller. 

\section{Turbulent transport barrier}

The collisionality $\nu_{\ast i}$ appearing in Eq.~(\ref{taustar}) usually increases with radius and is much higher at the edge of the plasma than in its core. As a result, the ion root is practically always realised in the plasma periphery. If there is an electron root in the core, there will thus be a transition region from positive and negative radial electric field at some intermediate radius. The width and structure of such a transition cannot be predicted from conventional local transport theory but can be calculated through global neoclassical transport simulations \citep{Matsuoka,Fu,Kuczynski}. The transition region is usually found to be much narrower than the minor radius, thus resulting in a strongly sheared ${\bf E} \times {\bf B}$ flow. 

It has been known for several decades that such flows tend to suppress, or substantially reduce, plasma turbulence \citep{Burrell,Terry,Diamond}. Such reduction occurs approximately when the shearing rate exceeds the largest linear microinstability growth rate \citep{Waltz,Ivanov}
	 \bn \frac{1}{B}\frac{d E_r}{dr} \gtrsim \gamma_{\rm max}. 
	\label{Waltz}
	\en
For electrostatic instabilities with wavelengths comparable to the ion gyroradius, the latter is equal to \citep{Helander-2022}
	$$ \gamma_{\rm max} = \frac{\alpha v_{Ti}}{L_\perp}, $$
where $L_\perp$ is the length scale of radial pressure variation, and $\alpha$ is a constant of the order of a few percent in typical tokamaks and stellarators \citep{Podavini}. If the width of the transition region is $w$ and the electric field on either side of this region is of order $T_i/(eL_\perp)$, the Waltz criterion (\ref{Waltz}) thus becomes
	$$ \frac{T_i}{eBL_\perp w} \gtrsim \frac{\alpha v_{Ti}}{L_\perp}, $$
i.e.,
	$$ w \lesssim \frac{\rho_i}{\alpha}. $$
If this condition is satisfied, turbulence is expected to be suppressed in the transition region, thus causing a steepening of the temperature profile there and a temperature increase in the plasma core of the order of 
	$$ \frac{\Delta T}{T} \sim w \left| \frac{d \ln T}{dr} \right| 
	\sim \frac{w}{L_\perp} \lesssim \frac{\rho_i}{\alpha L_\perp}. $$
Since $\alpha \ll 1$, this increase can be substantial for relevant values of $L_\perp$ (though not so in a reactor with $\rho_i / L_\perp \ll \alpha$),  but it is difficult to make a quantitative prediction since the temperature profile then steepens locally, which affects both the turbulence and the neoclassical transport. In any case, it seems likely that a transport barrier could arise at the electron-ion-root transition, potentially increasing core-plasma confinement. As already mentioned, steep electron-temperature profiles are indeed observed in electron-root plasmas \citep{Yamada_2005}.

\section{An explicit example}

\subsection{Magnetic configuration}

We now proceed to show a stellarator plasma which has been designed to have an electron root.
As we have seen, it is possible, in principle, to optimise for easy access to the electron root, but such a device will only be attractive as a power plant if it satisfies a range of other requirements too. 
For this reason, we adopt the optimisation strategy outlined by \citet{Goodman2024}, who devised a method by which a stellarator magnetic field can be numerically optimised to be quasi-isodynamic (QI) while simultaneously satisfying several other essential physics criteria. Here, we make two amendments to this method.
First, we add an additional target function, suggested by \cite{Kappel}, as an inequality constraint, to improve the configuration's compatibility with a set of practically feasible filamentary coils.
This additional target had little-to-no effect on the physics performance of the resulting configuration.

Second, and more importantly, we make several modifications to the implementation of the QI target function employed by \cite{Goodman2024}, which, in essence, penalises differences between the adiabatic invariant (\ref{eq:J}) corresponding to some input field, $\mathcal{J}_I$, and that in a constructed, perfectly QI field, $\mathcal{J}_C$. 
These differences are measured along various field lines on a flux surface, as well as multiple values of $y$. 
The form of this target is thus given by
\begin{equation}
    f_\textrm{QI} \propto \sum_{\tilde{y}} \sum_i \sum_j^{j_\textrm{max}} \left[ \mathcal{J}_I(\psi,\alpha_i,\tilde{y})^{p_\mathcal{J}} - \mathcal{J}_C(\psi, \alpha_j, \tilde{y})^{p_\mathcal{J}}\right]^2.
    \label{eq:fqinew}
\end{equation}
\noindent 
where $p_\mathcal{J} = 1$ and the various values of $y$, $\alpha_i$, and $\alpha_j$ are measured on uniform grids. 
Through numerical optimisation, $f_\textrm{QI}$ can be minimised on various flux-surfaces in an equilibrium, resulting in QI fields. 
In this work, we make three minor modifications to this target function.

First, since we need not necessarily sample $y$ on a uniform grid, we biased our optimisation in favour of more shallowly trapped particles by sampling more values of $y$ near $B_\textrm{max}$ than $B_\textrm{min}$.
The exact spacing of these points was varied throughout the optimisation to ensure that the confinement of deeply trapped particles was not spoiled too significantly.
We parameterise this spacing with the term $p_y$, and define the $i$th value of our new $y$ grid as
$$
\tilde{y}_i = B_\textrm{min} + \left(\frac{i}{N_y}(B_\textrm{max} - B_\textrm{min})^{1/p_y}\right)^{p_y}
$$
By increasing $p_y$, we create more points near $B_\textrm{min}$ than $B_\textrm{max}$. 
We varied $p_y$ in the range [0.5, 2] throughout this optimisation, as we felt needed.

Second, we need not necessarily compare $\mathcal{J}$ along every field line $\alpha_i$ to every other field line $\alpha_j$ on the same flux surface.
For instance, one could consider only comparing neighbouring field lines, thus more-closely approximating an expression for $\partial\mathcal{J}/\partial\alpha$.
We found that, by increasing the range of field lines along which we compared $\mathcal{J}$, shallowly trapped particles were better confined.
This range of field lines is parameterised in $j_\textrm{max}$ in Eq.~(\ref{eq:fqinew}), and was also varied throughout the optimisation process.
At times, $j_\textrm{max}$ was large enough to encompass entire flux surface, while at other times, it was small enough to see only neighbouring field lines.

Finally, we note that $\mathcal{J}$ is always larger for shallowly trapped particles than for deeply trapped ones, since both the integrand and the range of integration in Eq.~(\ref{eq:J}) is larger.
Hence, another tool to bias the optimisation towards better confined shallowly trapped particles is to increase the parameter $p_\mathcal{J}$ in Eq.~(\ref{eq:fqinew}), which increases the relative importance of these particles in the optimisation.
As with the other new parameters, $p_\mathcal{J}\in[0.5, 2]$ was varied throughout the optimisation, which was carried out with the \texttt{simsopt} optimisation framework \citep{simsopt} using the \texttt{VMEC} magnetohydrodynamic (MHD) equilibrium code \citep{VMEC}.

\begin{figure}
    \begin{center}
        \includegraphics[width=\linewidth]{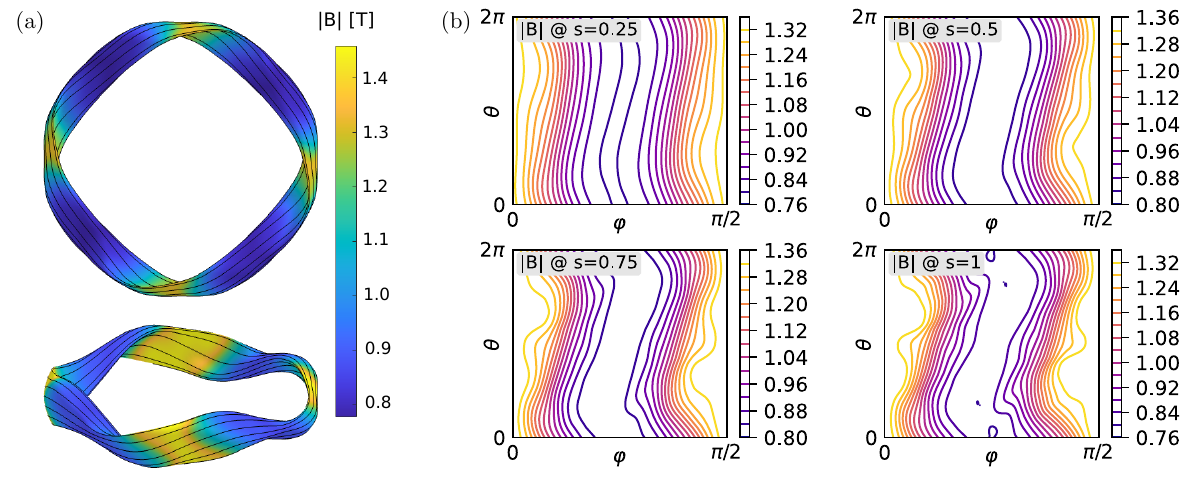}
        \caption{(left) the boundary of the configuration found in this work, with field lines shown in black; (right) this configuration's $B$-contours, in Boozer coordinates, on various flux surfaces within this configuration.} 
        \label{fig:squid}
    \end{center}
\end{figure}

The result of this optimisation is a QI stellarator configuration, shown in Fig.~\ref{fig:squid}, with many attractive properties, the majority of which will be discussed in a subsequent publication \citep{Goodman2024_2}. For now, we focus on its properties relevant to this work. The configuration has visually excellent QI quality, as evidenced by its poloidally closed $B$-contours shown in Fig.~\ref{fig:squid}. The coefficient $\epsilon_{\rm eff}$ varies from 0.004 on the magnetic axis to 0.025 at the plasma edge. We also note that, while the quality of the omnigenity of its deeply trapped particles is diminished, this configuration is able to confine the high-energy ions generated from fusion reactions as well as other highly optimised configurations \citep{Goodman-2023,Goodman2024,LandremanPaul}. In collisionless orbit-following simulations for 0.2s (which corresponds to a typical slowing-down time), no particles launched at half radius are lost, if the magnetic field is scaled to a mean on-axis field strength of $5.4\textrm{ T}$ and
the minor radius of the plasma exceeds 60 cm. 
This is due to the relatively large radial gradient of $\mathcal{J}$, which is known to improve the confinement of these ions \citep{Sanchez2023,Velasco_2023}.

For the purposes of the present paper, the most important property of the configuration is the possibility of an electron root for reactor-relevant plasma parameters, to which we now turn our attention. 

\subsection{Performance under reactor conditions}

To model reactor plasmas, the optimised equilibrium was scaled to  
a volume of 1450~m$^3$, yielding a major radius of $R=20.13$~m and a 
minor radius of $a=1.91$~m.  The average magnetic field strength at the
plasma axis was taken to be $B_0=5.4$~T.  Simulations were performed
for this equilibrium with the predictive version of the 1-D transport
code NTSS \citep{Turkin}.  A density scan was performed,
varying the central densities of deuterium and tritium, but without
changing the profile shape.  NTSS then solves the coupled energy balance
equations for electrons ($\sigma=e$), deuterium ($\sigma={\rm D}$),
tritium ($\sigma={\rm T}$) and helium ash ($\sigma={\rm He}$),
modelling the energy flux densities as the sum of neoclassical and
turbulent contributions, $Q_\sigma = Q_\sigma^{\rm neo} + Q_\sigma^{\rm turb}$.
The neoclassical portion is determined using tabulated values of the
mono-energetic transport transport coefficients calculated by the
Drift Kinetic Equation Solver (DKES) \citep{DKES}, while
the part from turbulence is modelled by the simple formula $Q_\sigma^{\rm turb}
= -n_\sigma \chi^{\rm turb} ({\rm d}T_\sigma/{\rm d}r)$, with the magnitude
of $\chi^{\rm turb}$ being adjusted during the course of the density scan to
yield fusion plasmas producing $P_\alpha = 600$~MW of alpha-particle
heating power.  The profile of the radial electric field is
simultaneously calculated, with the particle flux densities also
being modelled as a sum of neoclassical and turbulent contributions,
$\Gamma_\sigma = \Gamma_\sigma^{\rm neo} + \Gamma_\sigma^{\rm turb}$.  The
neoclassical portion can again be calculated from the DKES data set,
and the turbulent portion is obtained from $\Gamma_\sigma^{\rm turb}=
- D^{\rm turb} ({\rm d}n_\sigma/{\rm d}r)$, with the additional assumption
that $\chi^{\rm turb}/D^{\rm turb} = 7.5$.  A differential equation is solved for
the radial electric field to ensure a unique value of $E_r$ in parts
of the plasma where multiple roots of the ambipolarity constraint 
exist.  This equation serves to minimize the generalized heat production
rate, thereby introducing thermodynamic considerations into the
determination of the radius at which root transitions take place
\citep{Shaing,Turkin}.

\begin{figure}
    \centering
    \subfigure[Density profiles]{
    \label{fig:a}
        \includegraphics[width=0.3\textwidth]{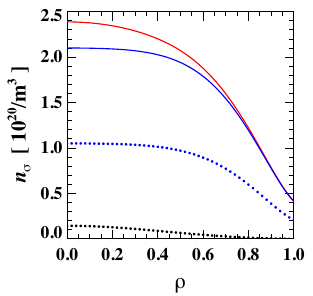}
        }
    \subfigure[Temperature profiles]{
        \label{fig:b}
        \includegraphics[width=0.3\textwidth]{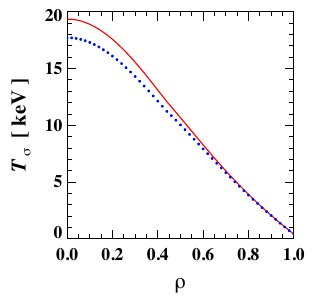}
        }
         \subfigure[Electric field profile]{
        \label{fig:c}
        \includegraphics[width=0.3\textwidth]{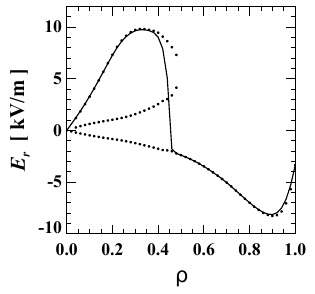}
        }
    \caption{{(a):} Assumed density profiles of electrons (red solid), D+T fuel ions (blue solid), D and T individually (blue dotted), and He (black dotted), as functions of normalised radius $\rho = r/a$. 
    {(b):} Temperature profiles for electrons (red solid) and all ions (blue dotted).
    {(c):} Radial electric field calculated through diffusive model (black solid) and the three roots of the ambipolarity equation (dotted). Note the rapid transition from the electron root for $\rho < 0.4$ to the ion root for $\rho > 0.45$.}
\end{figure}

An example from this density scan is illustrated by the results depicted
in Figure~2.  The profiles of density, temperature and radial electric 
field are shown for the case with central deuterium/tritium densities
of $1.05 \times 10^{20}$~m$^{-3}$.  These profiles for the bulk-ion
species are shown in the $n_\sigma$ plot by the blue dotted curve, and
their sum is given by the blue continuous curve.  The density of helium
ash is shown by the black dotted curve and the electron density profile
is plotted in red.  The same color scheme is used to depict the 
temperature profiles, although the differences in $T_\sigma$ for the
the ion species are too small to be discernable on the plot.  The
profile of the radial electric field obtained from the differential
equation is shown by the black continuous curve, and the electron-
unstable- and ion-root solutions of the ambipolarity constraint are
given by the black data points.  The volume average of the normalized
plasma pressure in this simulation has the value of 
$\langle\beta\rangle = 3.87$~\%.

For this example, the electron root extends nearly to a normalized
plasma radius of $\rho=r/a=0.5$.  This extent is increased (decreased)
if the simulation is performed at lower (higher) density and higher
(lower) temperature, as expected theoretically.  In this example, the energy confinement
time is $\tau_E=1.71$~s, which is only slightly above the
$\tau_E=1.65$~s predicted by the International Stellarator Scaling
ISS04 \citep{Yamada_2005} for the results of this simulation.

It is interesting to note that, although the electron root is realised for $0 \le \rho \le 0.45$ in this calculation, two other roots also exist as solutions to the ambipolarity equation in this region. These are indicated by dotted curves in Figure 2. In the centre of the plasma, the electron root is much stronger (has larger amplitude) but the weaker ion root may also be realised. The diffusive model used to calculate $E_r$ exhibits hysteresis, and the outcome thus depends on the pre-history of the system \citep{Hastings-1986}. The electron root is selected if the temperature of the electrons is initially chosen to be higher than that of the ions, as would be the case in a device with dominant electron heating.

\section{Discussion}

We have shown that it appears possible to design the magnetic field in a stellarator in such a way that the radial electric field becomes positive in the plasma core under reactor conditions. Although the density needs to be high and $T_e \simeq T_i$ in an ignited plasma, an electron root should be attainable if the magnetic field is tailored properly.  

In order to attain this goal, the neoclassical particle fluxes should not be made {\em too} small. For instance, an exactly omnigenous field would not meet the necessary requirements. The neoclassical electron transport must be able to compete with ion transport in the $\sqrt{\nu}$-regime, and this condition can be met if the confinement of shallowly trapped ions is made to be better than that of deeply trapped ones. This aim is not achieved by traditional optimisation of neoclassical transport that merely aims at reducing the coefficient $\epsilon_{\rm eff}$ controlling the $1/\nu$-transport of the electrons. It is considerably simpler to attain an electron root if $\epsilon_{\rm eff}$ is not extremely small, but fortunately it does not need to be so large that the resulting neoclassical thermal transport is of concern. Moreover, despite the fact that magnetic field is not omnigenous, fast-ion losses can be made small enough.\footnote{Improving the confinement of shallowly trapped particles also reduces the stochastic diffusion of transitioning particles, which is often the dominant transport mechanism of fast ions \citep{Beidler-2001}. } Other main goals of stellarator optimisation can also be met, and it thus appears possible to design a reactor with an electron root in the plasma core. 

Such a design would have two advantages over conventional ion-root stellarators. There should be no, or very little, accumulation of impurities by inward neoclassical transport, and a turbulent transport barrier is likely to arise at the radius where the electric field switches sign from positive to negative. If this transport barrier is strong enough, the energy confinement time of the plasma would be significantly enhanced. Much more work is necessary to assess the confinement improvement through turbulence simulations, which should properly account for the radial electric field created by the neoclassical transport. Ideally, such simulations should compute the turbulent and the neoclassical transport in an integrated way, so as to include any turbulent modifications of the electric field, which may well be important on the narrow length scale of the transport barrier. 

\section*{Acknowledgements}
The authors are grateful to Richard Nies for pointing out an error in the estimate of the transport-barrier width in an earlier version of the manuscript. 
This work was partly supported by a grant from the Simons Foundation (560651, PH). Views and opinions expressed are 
those of the authors only and do not necessarily reflect those of the European Union or the European Commission. Neither the European Union nor the European
Commission can be held responsible for them.

\bibliographystyle{jpp}
\bibliography{bibfile}

\end{document}